\documentstyle[12pt]{article}

\renewcommand{\thefootnote}{\fnsymbol{footnote}}

\newfont{\goth}{eufm10 scaled\magstep1}

\def\ie{\mbox{{\it i.e.} }}

\def\eg{\mbox{{\it e.g.} }}
\def\inv#1{{1\over #1}}
\def\abs#1{\left|#1\right|}
\def\acomm#1#2{\left\{#1,#2\right\}}
\def\comm#1#2{\left[#1,#2\right]}
\def\part{\partial}
\def\W{{\cal W}_3}
\def\A#1{\bar{\cal A}\left[#1\right]}
\def\B#1{\bar{\cal B}\left[#1\right]}
\def\C#1{\bar{\cal C}\left[#1\right]}
\def\D#1{\bar{\cal D}\left[#1\right]}
\def\E#1{\bar{\cal E}\left[#1\right]}
\def\F#1{{\cal F}\left[#1\right]}
\def\G#1{{\cal G}\left[#1\right]}
\def\H#1{{\cal H}\left[#1\right]}
\def\I#1{{\cal I}\left[#1\right]}

\def\oconf{\omega}
\def\cward#1{\Omega\left[#1\right]}
\def\owalg{\bar{\omega}}
\def\wward#1{\bar{\Omega}\left[#1\right]}
\def\cact#1{\delta S_0\left[#1\right]}
\def\wact#1{\bar{\delta}S_0\left[#1\right]}
\def\canom#1{\Delta_0\left[#1\right]}
\def\wanom#1{\bar{\Delta}_0\left[#1\right]}

\def\var#1{\delta_{#1}}
\def\conf#1{\var{\epsilon_{#1}}}
\def\walg#1{\bar{\delta}_{\lambda_{#1}}}
\def\meas{\int_{\Sigma}d^2 z\,}
\def\anom#1{\Delta\left[#1\right]}
\def\banom#1{\bar{\Delta}\left[#1\right]}

\begin{document}
\begin{titlepage}
\begin{center}

Centre de Physique Th\'eorique\footnote{Unit\'e Propre de Recherche 7061} -
CNRS - Luminy, Case 907\\
F-13288 Marseille Cedex 9 - France

\vspace{2 cm}

{\large \bf Generalized Wess-Zumino Consistency Conditions for Pure $\W$
Gravity Anomalies}
\
\vspace{.5 cm}

(to appear in Proceedings, ``$\cal W$-Algebras: Extended Conformal
Symmetries'', Centre de Physique Theorique, Marseille, France, 3-7 July
1995, R.\ Grimm, V.\ Ovsienko (eds.))

\vspace{1 cm}

\setcounter{footnote}{0}
\renewcommand{\thefootnote}{\arabic{footnote}}

{\bf Paul WATTS}\footnote{E-mail: {\it watts@cpt.univ-mrs.fr} -- WWW Home
Page: {\it http://cpt.univ-mrs.fr/~watts}}$^,$\footnote{Address after 15
October 1995: Department of Physics, University of Miami, P.O.\ Box 248046,
Coral Gables, FL 33124-8046, USA, E-mail: {\it watts@phyvax.ir.miami.edu}}

\vspace{2 cm}

{\bf Abstract}

\end{center}

\noindent General expressions for the anomalies appearing in pure $\W$
gravity are found by requiring that they satisfy a modified version of the
Wess-Zumino consistency conditions in which the Ward identities are treated
as nonvanishing quantities.

\bigskip

\noindent Keywords: $\cal W$ algebras, anomalies, Ward identities

\bigskip

\noindent PACS-95: 11.25.Hf 11.30.Ly

\bigskip

\noindent 8 September 1995

\bigskip

\noindent CPT-95/P.3237

\bigskip

\noindent Anonymous FTP or gopher: {\it cpt.univ-mrs.fr}

\end{titlepage}

\newpage
\setcounter{page}{1}
\renewcommand{\thepage}{\arabic{page}}

\section{Introduction}
\setcounter{equation}{0}

Quantum $\cal W$ algebras have received much attention in recent years, due
to their appearance both as infinite-dimensional symmetry algebras in
conformal field theories \cite{BS,dB,Hull}, and as finite-dimensional
algebras arising from constraining the Kirillov Poisson algebra associated
with a Lie algebra {\goth g} with the constraints associated to a
particular embedding of {\goth sl}(2) in {\goth g} treated by means of BRS
quantization \cite{dBT1,dBT2}.

A particular example of the former case was studied in detail in
\cite{OSSvN}, in which the quantum energy-momentum tensor $T$ and quantum
$\W$ current $W$ were coupled to classical sources $h$ and $b$
respectively.  The partition function depending on these sources was then
found, and since the transformation properties of the quantum currents and
sources were given, the anomalies for this theory of induced $\W$ gravity
were determined.  These anomalies depended on both the sources and the
effective currents, and in the classical (\ie $\abs{c}\rightarrow\infty$)
limit, the transformations of the sources and effective fields were found
to form a closed algebra, which the authors referred to as the ``pure $\W$
gravity'' algebra.  The closure of this algebra therefore gave both the
Ward identities of the theory as well as insuring that the anomalies
satisfied the Wess-Zumino consistency conditions (WZCCs).

In \cite{GGL}, a BRS differential algebra with an underlying gauge symmetry
was considered.  This was then constrained, in a manner akin to the second
of the two approaches described above, by decomposing the Lie algebra
{\goth g} associated with this symmetry with respect to a particular {\goth
sl}(2) embedding and then imposing constraints insuring the vanishing of
the field strength (\ie a zero-curvature condition).  For the case of the
principal embedding of {\goth sl}(2) into {\goth sl}(3), the resulting
relations between the BRS algebra elements due to these constraints gave
the pure $\W$ gravity transformation algebra and the Ward identities,
identical to the ones in \cite{OSSvN}.  The anomalies were then determined
by requiring them to be conformally covariant local functionals of ghost
number unity which were BRS-closed (up to an overall derivative); these
criteria guaranteed that the WZCCs would also be satisfied.

However, the sets of anomalies from these two treatments were {\em not} the
same.  This fact immediately begs the question, ``Why not?''  The goal of
this work is to try to answer this question by finding the most general
possible forms of the conformal and $\W$ anomalies of pure $\W$ gravity
satisfying a modified version of the WZCCs.  The approach presented here
will be purely formal, by treating pure $\W$ gravity as the abstract
algebra of transformations on a set of classical fields, without worrying
whether this algebra came from a conformal field theory or a BRS algebra.

\section{Pure $\W$ Gravity}
\setcounter{equation}{0}

\subsection{Pure $\W$ Gravity Algebra}

The pure $\W$ gravity algebra (using the notation of \cite{OSSvN}) consists
of two sets of transformations defined on four classical fields in a
2-dimensional space $\Sigma$ (with coordinates $(z,\bar{z})$): the
stress-energy tensor $u$, the metric $h$, the $\W$ current $v$, and the
$\W$ gauge field $b$.  The first set are the usual linear conformal
transformations; they close on the two sectors, \ie $(u,h)\mapsto (u,h)$
and $(v,b)\mapsto (v,b)$.  More specifically, with the infinitesimal
transformations parametrized by functions $\epsilon$, these take the form
\begin{eqnarray}
\conf{}u&=&\part^3\epsilon+\epsilon\part u+2u\part\epsilon,\nonumber\\
\conf{}h&=&\bar{\part}\epsilon+\epsilon\part h-h\part\epsilon,\nonumber\\
\conf{}v&=&\epsilon\part v +3v\part\epsilon,\nonumber\\
\conf{}b&=&\epsilon\part b-2b\part\epsilon ,
\end{eqnarray}

The $\W$ transformations (which are denoted with a barred $\delta$ to
distinguish them from the conformal ones) mix the two sectors nonlinearly,
in the following manner:
\begin{eqnarray}
\walg{}u&=&\inv{15}\lambda\part v+\inv{10}v\part\lambda,\nonumber\\
\walg{}h&=&\inv{15}\lambda\part^3b-\inv{10}\part\lambda\part^2b+\inv{10}
\part^2\lambda\part b-\inv{15}\part^3\lambda b+\frac{8}{15}u(\lambda\part
b-b\part\lambda),\nonumber\\
\walg{}v&=&\part^5\lambda+2\lambda\part^3u+9\part\lambda\part^2u+15\part^2
\lambda\part u+10\part^3\lambda u+16u(u\part\lambda+\lambda\part u),
\nonumber\\
\walg{}b&=&\bar{\part}\lambda+2\lambda\part h-h\part\lambda ,
\end{eqnarray}
where $\lambda$ is the variation parameter.

There are also two particular combinations of the fields, $\oconf$ and
$\owalg$, which play a major role in the study of pure $\W$ gravity
anomalies, given by
\begin{eqnarray}
\oconf&:=&\part^3h+2u\part h+h\part u+\inv{15}b\part v+\inv{10}v\part b-
\bar{\part}u,\nonumber\\
\owalg&:=&\part^5b+16u(u\part b+b\part u)+2b\part^3u+9\part b\part^2u+15
\part^2b\part u+10\part^3bu\nonumber\\
&&+3v\part h+h\part v-\bar{\part}v.
\end{eqnarray}
Under the conformal and $\W$ transformations, these transform as
\begin{eqnarray}
\conf{}\oconf&=&\epsilon\part\oconf+2\oconf\part\epsilon,\nonumber\\
\walg{}\oconf&=&\inv{15}\lambda\part\owalg+\inv{10}\owalg\part\lambda,
\nonumber\\
\conf{}\owalg&=&\epsilon\part\owalg+3\owalg\part\epsilon,\nonumber\\
\walg{}\owalg&=&2\lambda\part^3\oconf+9\part\lambda\part^2\oconf+15
\part^2\lambda\part\oconf+10\part^3\lambda\oconf\nonumber\\
&&+16u(\oconf\part\lambda+\lambda\part\oconf).\label{Ward-trans}
\end{eqnarray}

On the fields $u$, $v$ and $b$, it may be shown explicitly that the
transformations form a Lie algebra, with the commutators given by
\begin{eqnarray}
\comm{\conf{1}}{\conf{2}}=\var{\acomm{\epsilon_1}{\epsilon_2}},&
\comm{\conf{}}{\walg{}}=\bar{\delta}_{\acomm{\epsilon}{\lambda}},&
\comm{\walg{1}}{\walg{2}}=\var{\acomm{\lambda_1}{\lambda_2}},\label{Lie}
\end{eqnarray}
where the Poisson bracket $\acomm{\,}{\,}$ is defined on the space of
variations by
\begin{eqnarray}
\acomm{\epsilon_1}{\epsilon_2}&=&\epsilon_2\part\epsilon_1-\epsilon_1\part
\epsilon_2,\nonumber\\
\acomm{\epsilon}{\lambda}&=&2\lambda\part\epsilon-\epsilon\part\lambda,
\nonumber\\
\acomm{\lambda_1}{\lambda_2}&=&\inv{15}\lambda_2\part^3\lambda_1-\inv{10}
\part\lambda_2\part^2\lambda_1+\inv{10}\part^2\lambda_2\part\lambda_1-
\inv{15}\part^3\lambda_2\lambda_1\nonumber\\
&&+\acomm{\lambda_1}{\lambda_2}',\label{Poisson}
\end{eqnarray}
where
\begin{equation}
\acomm{\lambda_1}{\lambda_2}'=\frac{8}{15}u(\lambda_2\part\lambda_1
-\lambda_1\part\lambda_2).
\end{equation}
(The reason for splitting the last Poisson bracket up in this way will
become apparent in Section \ref{sec-basis}.)  However, on $h$, although the
first two of (\ref{Lie}) hold, the third takes the form
\begin{equation}
\comm{\walg{1}}{\walg{2}}h-\delta_{\acomm{\lambda_1}{\lambda_2}}h=
\frac{8}{15}\left(\lambda_2\part\lambda_1-\lambda_1\part\lambda_2\right)
\oconf,
\end{equation}
so, {\it a priori}, the algebra of transformations does {\em not} close.

\subsection{Ward Identities and Equivalence Relations}

Recall that in \cite{OSSvN}, this algebra was obtained as the classical
limit of an effective field theory coming from a quantum action which
exhibited conformal and $\W$ symmetries.  In this context, the algebra must
close, and this gives the Ward identities of the theory, namely the
vanishing of $\oconf$ and $\owalg$ (the latter condition arising from
consistency with the second and fourth of (\ref{Ward-trans})).  In
\cite{GGL}, the same algebra and Ward identities came about as consequences
of the zero-curvature conditions constraining the BRS algebra.

However, if the situation is treated purely formally, then one cannot use
either of these arguments to eliminate the extraneous term in the $\W-\W$
transformation of $h$.  Instead, this is accomplished by realizing that
because $\oconf$ and $\owalg$ transform into one another, any quantity of
the form $A+f(\oconf)$, where $f$ is some linear function, will change
under conformal transformations as $\conf{}A+f'(\oconf)$ and under $\W$
transformations as $\walg{}A+f''(\owalg)$, where $f'$ and $f''$ are some
other linear functions, and similarly for $\owalg$.  Therefore, one may
consistently treat two quantities as equivalent if they differ by some
linear function of $\oconf$ and/or $\owalg$.  Thus, by adopting this point
of view, one sees that the algebra of transformations does in fact close,
even on $h$.

\section{Generalized Wess-Zumino Consistency Conditions}

In order to extend the approach outlined in the previous section to
anomalies rather than classical fields, one must introduce the functionals
$\Omega$ and $\bar{\Omega}$ given by
\begin{eqnarray}
\cward{\epsilon}&\equiv&30\meas\epsilon\,\oconf,\nonumber\\
\wward{\lambda}&\equiv&\meas\lambda\,\owalg.
\end{eqnarray}
The transformation laws (\ref{Ward-trans}) imply that
\begin{eqnarray}
\conf{1}\cward{\epsilon_2}=\cward{\acomm{\epsilon_1}{\epsilon_2}},&&\walg{}
\cward{\epsilon}=\wward{\acomm{\lambda}{\epsilon}},\nonumber\\
\conf{}\wward{\lambda}=\wward{\acomm{\epsilon}{\lambda}},&&\walg{1}\wward{
\lambda_2}=\cward{\acomm{\lambda_1}{\lambda_2}}.\label{omega}
\end{eqnarray}

These two functionals are linear both in their arguments and in $\oconf$
and $\owalg$ respectively, and transform into one another.  Therefore, two
functionals $A$ and $B$ can be considered equivalent if they differ by
$\Omega$ and/or $\bar{\Omega}$, where the latter two may have {\em any}
arguments whatsoever.  This is entirely consistent; if any objects linear
in $\oconf$ and $\owalg$ are going to be modded away, it doesn't matter
what's multiplying them inside the integrals.

Recall the usual form of the WZCCs: if one has a closed algebra of
variations $\delta_{\xi}$, where $\xi$ is the variation parameter, then an
anomaly $\anom{\xi}$ is a linear functional of $\xi$ which must satisfy
\begin{equation}
\delta_{\xi_1}\anom{\xi_2}-\delta_{\xi_2}\anom{\xi_1}=\anom{\acomm{\xi_1}{
\xi_2}}.
\end{equation}

In the case considered here, there are two such variations, but the algebra
does not close.  However, the existence of the functionals $\cward{
\epsilon}$ and $\wward{\lambda}$ allow the following generalization of the
WZCCs which will rectify this: let $\anom{\epsilon}$ and $\banom{\lambda}$
be functionals of the conformal and $\W$ parameters respectively.  They
will be considered to be anomalies of the theory if and only if they
satisfy the following three conditions:
\begin{eqnarray}
\conf{1}\anom{\epsilon_2}-\conf{2}\anom{\epsilon_1}&=&\anom{\acomm{
\epsilon_1}{\epsilon_2}}+\cward{f_1(\epsilon_1,\epsilon_2)},\nonumber\\
\walg{1}\banom{\lambda_2}-\walg{2}\banom{\lambda_1}&=&\anom{\acomm{
\lambda_1}{\lambda_2}}+\cward{f_2(\lambda_1,\lambda_2)},\nonumber\\
\conf{}\banom{\lambda}-\walg{}\anom{\epsilon}&=&\banom{\acomm{\epsilon}{
\lambda}}+\wward{f_3(\epsilon,\lambda)},\label{WZCCs}
\end{eqnarray}
where $f_{1,2,3}$ are {\em any} functions of the appropriate variation
parameters (not necessarily the Poisson brackets given by (\ref{Poisson})).

\section{Pure $\W$ Gravity Anomalies with Generalized
WZCCs}\label{sec-basis}
\setcounter{equation}{0}

\subsection{A Basis for Pure $\W$ Gravity Anomalies}

One can define the following five quantities, which are functionals of the
$\W$ variation parameter $\lambda$:
\begin{eqnarray}
\A{\lambda}&:=&-\meas\lambda\part^5b,\nonumber\\
\B{\lambda}&:=&16\meas\lambda u(u\part b+b\part u),\nonumber\\
\C{\lambda}&:=&\meas\lambda(2b\part^3u+9\part b\part^2u+15\part^2b\part
u+10\part^3bu),\nonumber\\
\D{\lambda}&:=&\meas\lambda(3v\part h+h\part v),\nonumber\\
\E{\lambda}&:=&\meas\lambda\bar{\part}v;\label{wfuncts}
\end{eqnarray}
and the four functionals of the conformal parameter $\epsilon$:
\begin{eqnarray}
\F{\epsilon}&:=&-30\meas\epsilon\part^3h,\nonumber\\
\G{\epsilon}&:=&\meas\epsilon(2u\part h+h\part u),\nonumber\\
\H{\epsilon}&:=&\inv{30}\meas\epsilon(2b\part v+3v\part b),\nonumber\\
\I{\epsilon}&:=&\meas\epsilon\bar{\part}u.\label{cfuncts}
\end{eqnarray}
However, note that
\begin{eqnarray}
\cward{\epsilon}&:=&-\F{\epsilon}+30\left(\G{\epsilon}+\H{\epsilon}\right)
-30\I{\epsilon},\nonumber\\
\wward{\lambda}&:=&-\A{\lambda}+\B{\lambda}+\C{\lambda}+\D{\lambda}-
\E{\lambda},
\end{eqnarray}
so that any two of the just-defined functionals can be replaced by
$\wward{\lambda}$ and $\cward{\epsilon}$.  The choices made here are to get
rid of $\C{\lambda}$ and $\H{\epsilon}$.

The transformation properties for $\cward{\epsilon}$ and $\wward{\lambda}$
have already been given in (\ref{omega}); for the $\W$ variations of the
above functionals, one obtains
\begin{eqnarray}
\walg{1}\A{\lambda_2}-\walg{2}\A{\lambda_1}&=&\F{\acomm{\lambda_1}{
\lambda_2}}-\F{\acomm{\lambda_1}{\lambda_2}'},\nonumber\\
\walg{1}\B{\lambda_2}-\walg{2}\B{\lambda_1}&=&\F{\acomm{\lambda_1}{
\lambda_2}'}+\cward{\acomm{\lambda_1}{\lambda_2}'},\nonumber\\
\walg{1}\E{\lambda_2}-\walg{2}\E{\lambda_1}&=&30\I{\acomm{\lambda_1}{
\lambda_2}},
\end{eqnarray}
and for the conformal transformations,
\begin{eqnarray}
\conf{1}\F{\epsilon_2}-\conf{2}\F{\epsilon_1}&=&\F{\acomm{\epsilon_1}{
\epsilon_2}},\nonumber\\
\conf{1}\I{\epsilon_2}-\conf{2}\I{\epsilon_1}&=&\I{\acomm{\epsilon_1}{
\epsilon_2}},\label{conf-anom}
\end{eqnarray}
and finally, the $\W$ variations of the conformal functionals are
\begin{eqnarray}
\walg{}\F{\epsilon}&=&\conf{}\A{\lambda}+\conf{}\B{\lambda}+\A{\acomm{
\lambda}{\epsilon}}+\B{\acomm{\lambda}{\epsilon}},\nonumber\\
\walg{}\I{\epsilon}&=&\inv{30}\left(\conf{}\E{\lambda}+\E{\acomm{\lambda}{
\epsilon}}\right).\label{mix}
\end{eqnarray}
(The reason that $\D{\lambda}$ and $\G{\epsilon}$ have disappeared
completely comes from the fact that they always appears in the combinations
$\C{\lambda}+\D{\lambda}$ and $\G{\epsilon}+\H{\epsilon}$, so eliminating
$\C{\lambda}$ and $\H{\epsilon}$ in favor of $\wward{\lambda}$ and
$\cward{\epsilon}$ respectively eliminates $\D{\lambda}$ and $\G{\epsilon}$
as well.)

Notice that the only combinations appearing in the mixed transformations
(\ref{omega}) and (\ref{mix}) are $\F{\epsilon}$, $\I{\epsilon}$,
$\cward{\epsilon}$, $\A{\lambda}+\B{\lambda}$, $\E{\lambda}$, and
$\wward{\lambda}$.  So, the last of (\ref{WZCCs}) implies that these are
the only possible terms which can appear in $\Delta$ and $\bar{\Delta}$.
With this in mind, it now becomes more convenient to change bases to the
three conformal functionals $\cward{\epsilon}$, $\canom{\epsilon}$ and
$\cact{\epsilon}$, and the three $\W$ functionals $\wward{\lambda}$,
$\wanom{\lambda}$ and $\wact{\lambda}$, where
\begin{eqnarray}
\canom{\epsilon}&:=&\F{\epsilon},\nonumber\\
\cact{\epsilon}&:=&\F{\epsilon}-30\I{\epsilon},\nonumber\\
\wanom{\lambda}&:=&\A{\lambda}+\B{\lambda},\nonumber\\
\wact{\lambda}&:=&\A{\lambda}+\B{\lambda}-\E{\lambda}.\label{basis}
\end{eqnarray}
Not only do these choices simplify the basis, but notice that
$\cact{\epsilon}$ and $\wact{\lambda}$ are precisely what the notations
suggest, \ie the appropriate variations of the ``action''
\begin{equation}
S_0:=\meas\left(30uh+vb\right).
\end{equation}

\subsection{Pure $\W$ Gravity Anomalies}

The transformations of the new basis elements (\ref{basis}) are
\begin{eqnarray}
\conf{1}\canom{\epsilon_2}-\conf{2}\canom{\epsilon_1}&=&\canom{\acomm{
\epsilon_1}{\epsilon_2}},\nonumber\\
\walg{1}\wanom{\lambda_2}-\walg{2}\wanom{\lambda_1}&=&\canom{\acomm{
\lambda_1}{\lambda_2}}+\cward{\acomm{\lambda_1}{\lambda_2}'},
\nonumber\\
\conf{}\wanom{\lambda}-\walg{}\canom{\epsilon}&=&\wanom{\acomm{\epsilon}{
\lambda}},
\end{eqnarray}
and
\begin{eqnarray}
\conf{1}\cact{\epsilon_2}-\conf{2}\cact{\epsilon_1}&=&\cact{\acomm{
\epsilon_1}{\epsilon_2}},\nonumber\\
\walg{1}\wact{\lambda_2}-\walg{2}\wact{\lambda_1}&=&\cact{\acomm{\lambda_1
}{\lambda_1}}+\cward{\acomm{\lambda_1}{\lambda_2}'},\nonumber\\
\conf{}\wact{\lambda}-\walg{}\cact{\epsilon}&=&\wact{\acomm{\epsilon}{
\lambda}}.
\end{eqnarray}

Thus, under the generalized WZCCs, all six of the basis elements are
permissible pure $\W$ gravity anomalies, so the most general possible
conformal and $\W$ anomalies will just be linear combinations of these
functionals, subject to (\ref{WZCCs}); it is easy to show that these must
have the form
\begin{eqnarray}
\anom{\epsilon}&=&\alpha\canom{\epsilon}+\beta\cact{\epsilon}+\gamma
\cward{\epsilon},\nonumber\\
\banom{\lambda}&=&\alpha\wanom{\lambda}+\beta\wact{\lambda}+\tau\wward{
\lambda},
\label{general}
\end{eqnarray}
for arbitrary constants $\alpha$, $\beta$, $\gamma$ and $\tau$.

\subsection{Examples}

In order to illustrate the utility of the approach of this work, the two
sets of anomalies from \cite{OSSvN} and \cite{GGL} are examined and shown
to both come from the general form (\ref{general}) with specific values of
the constants $\alpha$, $\beta$, $\gamma$, and $\tau$.

The comparison with the anomalies in \cite{OSSvN} is easy, since the
notation used herein is the same; they correspond to the constants above
taking the particular values $\alpha=\frac{c}{360\pi}$, $\beta=\gamma=\tau
=0$, and thus are of the form
\begin{eqnarray}
\anom{\epsilon}&=&\frac{c}{360\pi}\canom{\epsilon},
\nonumber\\
\banom{\lambda}&=&\frac{c}{360\pi}\wanom{\lambda}.
\end{eqnarray}
{}From a field-theoretical point of view, these are in fact the only possible
anomalies; they both come from the variations of an effective action
arising from a quantum action which is invariant under the $\W$ algebra.
Thus, there are no pieces involving $\oconf$ and $\owalg$ (due to the Ward
identities), and automatically satisfy the usual WZCCs (since the algebra
closes).

To consider the anomalies obtained in \cite{GGL}, one must first find the
relations between the fields $(\Lambda_{zz},v_{\bar{z}}{}^z,W_3,v_{
\bar{z}z}{}^z)$ and the BRS ghosts $(c^z,c^{\bar{z}z})$, and the fields
$(u,h,v,b)$ and the variation parameters $(\epsilon,\lambda)$; the correct
identifications are
\begin{eqnarray}
\Lambda_{zz}\mapsto u,&&v_{\bar{z}}{}^z\mapsto h,\nonumber\\
W_3\mapsto\frac{i}{24\sqrt{10}}v,&&v_{\bar{z}z}{}^z\mapsto\frac{i}{
\sqrt{10}}b,\nonumber\\
c^z\mapsto\epsilon,&&c^{\bar{z}z}\mapsto\frac{i}{\sqrt{10}}\lambda,
\end{eqnarray}
so that the anomalies take the form
\begin{eqnarray}
\anom{\epsilon}&=&\inv{15}\left(\canom{\epsilon}-\cact{\epsilon}+\cward{
\epsilon}\right),\nonumber\\
\banom{\lambda}&=&\inv{15}\left(\wanom{\lambda}-\wact{\lambda}+\wward{
\lambda}\right),
\end{eqnarray}
which correspond to $\alpha=-\beta=\gamma=\tau=\inv{15}$.  In this BRS
approach, the algebra closes and the Ward identities are satisfied, so in
reality, the anomalies in \cite{OSSvN} and \cite{GGL} differ only by the
appropriate variation of the quantity $S_0$.  This of course is not a
problem, because if the algebra closes, one can add {\em any} total
variations $\conf{}F$ and $\walg{}F$ of some field-dependent quantity $F$
to the respective anomalies, and the usual WZCCs will still be satisfied.

\section{Conclusion}
\setcounter{equation}{0}

By extending the WZCCs in the manner just presented, it is possible to
determine general forms for the anomalies appearing in pure $\W$ gravity,
and that two of the known cases do indeed have these forms.  However, there
is no reason that this same argument would not apply to other cases, \eg a
theory with a ${\cal W}_N$ symmetry; for the symmetry algebra of such a
theory to close, the quantities corresponding to $\omega$ and
$\bar{\omega}$ will transform into themselves, so that all Ward identities
hold covariantly.  This means that one can use the same trick, namely,
treat any two expressions differing by a linear combination of these
quantities as equivalent.  General forms of the anomalies may then be
obtained.

\section*{Acknowledgements}

I would like to thank Richard Grimm for bringing this subject to my
attention, and Daniela G\v{a}r\v{a}jeu for helpful discussions.

\newpage

\end{document}